\documentclass[a4paper]{jpconf}
\usepackage{graphicx}
\begin{document}
\title{Introduction to Galactic Chemical Evolution}

\author{Francesca Matteucci}

\address{Department of Physics, Trieste University, Via G.B. Tiepolo 11, 34131 Triestwe, Italy}

\ead{matteucci@oats.inaf.it}

\begin{abstract}
In this lecture I will introduce the concept of galactic chemical evolution, namely the study of how and where the chemical elements formed and how they were distributed in the stars and gas in galaxies. The main ingredients to build models of galactic chemical evolution will be described. They include: initial conditions, star formation history, stellar nucleosynthesis and gas flows in and out of galaxies. Then some simple analytical models and their solutions will be discussed together with the main criticisms associated to them. The yield per stellar generation will be defined and the hypothesis of instantaneous recycling approximation will be critically discussed. Detailed numerical models of chemical evolution of galaxies of different morphological type, able to follow the time evolution of the abundances of single elements, will be discussed and their predictions will be compared to observational data.  The comparisons will include stellar abundances as well as interstellar medium ones, measured in galaxies. I will show how, from these comparisons,  one can derive important constraints on stellar nucleosynthesis and galaxy formation mechanisms. Most of the concepts described in this lecture can be found in the monograph by Matteucci (2012).

\end{abstract}

\section{Introduction}
We call chemical evolution of galaxies the study of how the chemical elements formed in stars and were distributed in galaxies. During the Big Bang only the light elements (H, D, He, Li) were synthesized, while stars have been responsible for the formation and distribution of all the elements from carbon to uranium and beyond. Some light elements, such as Li, Be and B are formed during the spallation process, which is the interaction between the cosmic rays and C,N,O atoms present in the interstellar medium (ISM). Therefore, stars produce chemical elements in their interiors by means of nuclear reactions of fusion and then restore these elements into the ISM when they die. The fusion reactions occur up to $^{56}Fe$ which is the element with the maximum binding energy per nucleon. For elements heavier than  $^{56}Fe$ the nuclear fusion is therefore inhibited and the nuclear fission is favored. The stars are born, live and die and they can die in a quiescent fashion like white dwarfs or violently as supernovae (SNe). The supernovae can be of various types: core-collapse SNe, namely the explosion of single massive stars ($M \ge 10 M_{\odot}$, Type II, Ib, Ic) , and  Type Ia SNe, namely the explosion of a white dwarf occurring after accreting material from a companion in a binary system.
In this lecture I will first describe the ingredients necessary to build galactic chemical evolution models. Such models can be analytical or numerical and they aim at following the evolution in time and space of the abundances of the chemical elements in the ISM. Then,  I will focuse on some highlights in the galactic chemical evolution and on the comparison models-observations. In doing so, I will show how chemical evolution models can constrain stellar nucleosynthesis and galaxy formation timescales.

\section{Basic ingredients for galactic chemical evolution}
The basic ingredients necessary to build a chemical evolution model are: 
\begin{itemize}
\item {\bf Initial conditions}

The initial conditions consist in deciding whether to assume an open or a closed model, in other words to decide if the gas, out of which the stars 
form, is present since the beginning or if it is accreted during the galactic lifetime; moreover, one should assume whether the initial gas has a primordial (only light elements from the Big Bang) or enriched chemical composition.

\item {\bf Stellar birthrate function (SFRxIMF)}

The stellar birthrate function is the history of star formation in a galaxy and it can be expressed as the product of the star formation rate (SFR) per the initial mass function (IMF), namely:

\begin{equation}
B(m,t)= \psi(t) \cdot \phi(m),
\end{equation}
where the function $\psi(t)$ represents the SFR and it is generally assumed to be only a function of time, whereas $\phi(m)$ is the IMF which is assumed to be only a function of mass. The SFR represents how many solar masses go into stars per unit time, while the IMF is the initial stellar mass function describing the distribution of stars at birth as a function of the stellar mass. Clearly, these hypotheses are semplifications and we do not know whether the SFR is independent of mass and the IMF independent of time.

\subsection{Parametrization of the SFR}

The most common parametrization is the Schmidt (1959) law where the SFR is proportional to some power $k$ of the gas density. Kennicutt (1998) suggested $k=1.4 \pm 0.15$, as deduced by data relative to  star forming galaxies.
Other important parameters such as gas temperature, viscosity and magnetic field are usually ignored.

The Kennicutt SFR can be written as:
\begin{equation}
\psi(t)= \nu \sigma_{gas}^{k},
\end{equation}
where $\sigma_{gas}^{k}$ is the gas surface mass density and $\nu$ is the efficiency of star formation, namely the SFR per unit mass of gas.


\begin{figure}[h]
\includegraphics[width=14pc]{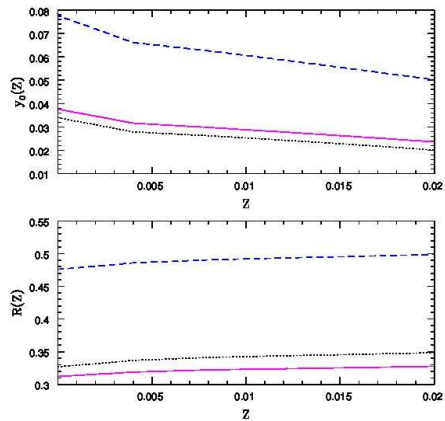}\hspace{2pc}%
\begin{minipage}[b]{14pc}\caption{Upper panel: the yield  of oxygen per stellar generation  computed for different metallicities and IMFs
The blue dotted line refers to Chabrier (2003) IMF, the magenta line is the Salpeter (1955) IMF and the black line is the Kroupa et al. (1993) IMF. Lower panel:returned fraction R as a function of IMF and metallicity. Figure from Vincenzo et al. (2015).}
\end{minipage}
\end{figure}

\subsection{Parametrization of the IMF}
The IMF, namely the number of stars born in the mass interval, m-m+dm, is generally expressed as a power law. It is possible to measure the IMF only in the solar vicinity since one needs to count the stars as functions of their magnitudes and it is still not possible to do it in external galaxies. Therefore, the only observational information we have is relative to the solar region of our Galaxy.

\item {\bf Stellar nucleosynthesis}

The stellar yields are defined as the masses of chemical elements produced by stars of different masses. These yields can represent both the newly formed elements or the elements already present in the star at its formation and restored into the ISM without being reprocessed. In particular, we can define the yield of a newly formed element, $p_{im}$, as the integral over the stellar lifetime of the mass loss rate multiplied by   [X(i)-Xo(i)], where Xo(i) is the original abundance of the element $i$. To obtain the total stellar yield, it should be added to this term the mass ejected without processing, namely  Xo(i) x mass lost, where with mass lost we intend the total mass ejected by a star into the ISM during its lifetime.
Each stellar mass can produce and eject different chemical elements and the yields are therefore a function of the stellar mass but also of the original stellar metal content that we will indicate with $Z$. These yields are computed by means of detailed nucleosynthesis calculations taking into account all the main nuclear reactions in stars.
Here we summarize briefly the element production in stars: i) low and intermediate mass stars (0.8-8 $M_{\odot}$): produce He, N, C and heavy s-process elements. They die as C-O white dwarfs, when single, and can die as Type Ia SNe when binaries.
ii) Stars with $M<0.8M_{\odot}$ do not contribute to galactic chemical enrichment and have lifetimes longer than the Hubble time.
iii) Massive stars ($M>8-10 M_{\odot}$): they produce mainly alpha-elements (O, Ne, Mg, S, Si, Ca),  some Fe, light s-process elements and perhaps r-process elements (?) and explode as core-collapse SNe. However, r-process elements originating in neutron binary mergers seems to represent the most promising channel for r-process element production these days (see Matteucci et al, 2014 and references therein).

\item {\bf Gas flows: infall, outflow, radial flows}

In order to build a realistic galaxy one has to assume the presence of gas flows both in and out. The gas inflows are considered either as gas accretion or radial gas flows and they are influencing the chemical evolution of galaxies: in the case of accretion , usually assumed at a constant rate or exponentially decreasing in time, the main effect is to dilute the metal content. In the case of outflows or galactic winds also the effect is to decrease the metal concentration by decreasing the gas which is available for star formation. Galactic outflows are generally assumed to occur at a rate proportional to the star formation rate.
\end{itemize}

\section{The yield per stellar generation}
The ``yield per stellar generation'' of a single chemical element, can be defined as (Tinsley 1980):

\begin{equation}
y_i={\int^{\infty}_{1}{mp_{im} \phi(m)dm} \over (1-R)}
\end{equation}
where $p_{im}$ is the mass of the newly produced element $i$ ejected by a star of mass $m$, and $R$ is the returned fraction. The yield $y_i$ is therefore the mass fraction of the element $i$ newly produced by a generation of stars relative to the fraction of mass in remnants (white dwarfs, neutron stars and black holes) and never dying low mass stars ($M< 0.1M_{\odot}$). 

We define ``returned fraction'' the fraction of mass ejected into the ISM by an entire stellar generation, namely:
\begin{equation}
R= \int^{\infty}_{1}{(m-M_{rem}) \phi(m) dm}
\end{equation}

The term fraction originates from the fact that both $y_i$ and $R$ are divided by the normalization condition of the IMF, namely:
\begin{equation}
\int^{\infty}_{0.1}{m \phi(m)dm}=1.
\end{equation}

In order to define $y_i$ and $R$
we have made a very specific assumption: the instantaneous recycling approximation (IRA), which states that {\it all stars more massive than 1$M_{\odot}$ die instantaneously, while all stars less massive live forever}. This assumption allows us to solve the chemical evolution equations analytically, as we will see in the following, but it is a very poor approximation for chemical elements produced on long timescales such as C, N and Fe. On the other hand, for oxygen, which is almost entirely produced by short lived core-collapse SNe, IRA is an acceptable approximation.
In Figure 1 we show $y_O$ and $R$,  computed for different initial metallicities of the stars and different IMFs. As one can see, the dependence of these quantities on Z is negligible whereas that on the IMF is strong. In that Figure it does not appear the Kroupa(2001) universal IMF, suggesting that the IMF in stellar clusters is an universal one. Kroupa (2001) IMF is a
two-slope IMF with a slope for stars more massive than 0.5$M_{\odot}$ very similar to that of the Salpeter (1955) IMF, still widely used in model for external galaxies.
\section{The Simple Model}
The Simple Model of chemical evolution assumes that the system is evolving as a closed-box, without inflows or outflows, the IMF is constant in time, the chemical composition of the gas is primordial and the mixing between the chemical products ejected by stars and the ISM is instantaneous, plus IRA.

The initial gas mass is therefore, $M_{gas}(0)=M_{tot}=constant$,  $\mu=M_{gas}/M_{tot}$ is the fractionary mass of gas, the metallicity is $Z=M_{Z}/M_{gas}$ is zero at the time t=0.
The basic equations can be written as:
\begin{equation}
{dM_{gas} \over dt}= -\psi(t)\,\,+  \int^{\infty}_{m(t)}{(m-M_{R}) \psi(t-\tau_m) \varphi(m) dm}
\end{equation}
which describes the evolution of gas and where the integral is the rate at which dying stars restore both the enriched and unenriched material into the ISM at the time $t$.

The equation for metals is:

\begin{equation}
{d(ZM_{gas}) \over dt}=-Z\psi(t) \,\,+ \int^{\infty}_{m(t)}{[(m-M_{R})Z(t-\tau_m)+
mp_{Zm}] \psi(t-\tau_m) \varphi(m)dm},  
\end{equation}
where the first term in the square brackets represents the mass of 
pristine metals which are restored into the ISM without suffering any 
nuclear processing, whereas the second term contains the newly formed 
and ejected metals (Maeder 1992).
When IRA is assumed, the SFR can be taken out of the integrals and the equation
for metals be  solved analytically. Its solution is:
\begin{equation}
Z= y_{Z} ln({ 1 \over \mu}). 
\end{equation}

The metallicity yield per stellar generation $y_Z$ which appears in the above 
equation is known as 
{\it effective yield}, simply defined as the yield $y_{Z_{eff}}$ that would be 
deduced if the system were assumed to be described by the Simple Model.

Therefore, the effective yield is:
\begin{equation}
y_{Z_{eff}}={Z \over ln(1/\mu)}. 
\end{equation}

Clearly, the true yield  $y_Z$ will be always lower than the effective one 
in both 
cases of winds and infall of primordial gas. The only way to increase the 
effective yield is to assume an IMF more weighted towards massive stars than 
the canonical IMFs.

\subsection{Primary and secondary elements}

We define primary element an element produced directly from H and He
A typical primary element is carbon or oxygen which originate from the 
3-$\alpha$ reactions
We define secondary element an element produced starting from metals already 
present in the star at birth (e.g. nitrogen produced in the CNO cycle).
We recall that the solution of the chemical evolution equations for a 
secondary element, with abundance $X_S$, implies:
\begin{equation}
X_S={ 1 \over 2}({y_S \over y_ZZ_{\odot}})Z^{2} ,
\end{equation}
where $y_S$ is the yield per stellar generation of the secondary element.
This means that, for a secondary element the simple closed-box model predicts
that its abundance increases proportionally to the metallicity squared, namely:

\begin{equation}
X_S \propto Z^{2} .
\end{equation}

\section{Analytical solutions for outflow and inflow}
In the case of a model with outflow but no inflow  occurring at a rate:
 \begin{equation}
W(t)=\lambda(1-R) \psi(t), 
\end{equation}
where $\lambda$ is a free parameter larger than or equal to zero, the solution
for the metallicity of the system is:

\begin{equation}
Z={y_Z  \over (1+ \lambda)}ln[(1 +\lambda) \mu^{-1}- \lambda]. 
\end{equation}
It is clear that in the case of $\lambda=0$ the solution is the same as that of the Simple Model.

In the case of a model without outflow but inflow of primordial gas 
($Z_{inf}=0$), occurring at a rate:
\begin{equation}
A(t)=\Lambda (1-R) \psi(t)
\end{equation}
with $\Lambda$ a positive constant different from zero and from 1, the solution is:
\begin{equation}
Z= {y_Z \over \Lambda}[1-(\Lambda-(\Lambda-1)\mu^{-1})^{-\Lambda/(1-\Lambda)}].
\end{equation}
Again, if $\Lambda=0$,  the solution is the same as that of the Simple Model.
The case $\Lambda=1$ is a particular one, called ``extreme infall'' and it 
has a different solution:
\begin{equation}
Z = y_Z (1-e^{- \beta}),
\end{equation}
 where $\beta=\mu^{-1}-1$.
For a more extensive discussion of these and other analytical solutions see Matteucci (2012).

\section{Numerical models} 
When the stellar lifetimes are correctly taken into account the chemical 
evolution equations should be solved numerically.
This allows us to follow in detail the temporal evolution of the abundances of single elements.
 
A complete chemical evolution model in the presence of both
galactic wind, gas infall
 and radial flows can be described by a number of equations equal to the number
of chemical species: in particular, if $M_i$ is the mass 
of the gas in the form of any chemical element $i$,
we can write the following set of integer-differential equations 
which  can be solved only numerically, if IRA is relaxed:

\begin{eqnarray}
\dot M_i(t) = -\psi(t)X_i(t)+
\int_{M_{L}}^{M_{Bm}}{\psi(t-\tau_m)
Q_{mi}(t-\tau_m)\varphi(m)dm}+\\ \nonumber
A_B\int_{M_{Bm}}^{M_{BM}}{\varphi(m)} \bigl[\int_{\mu_{B_{min}}}
^{0.5}{f(\mu_B)\psi(t-\tau_{m2}) Q_{mi}(t-\tau_{m2})d\mu_B \bigr]
dm}+\\ \nonumber (1-A_B)\int_{M_{Bm}}^
{M_{BM}}{\psi(t-\tau_{m})Q_{mi}(t-\tau_{m)}\varphi(m)dm}+\\
\nonumber \int_{M_{BM}}^{M_U}{\psi(t-\tau_m)Q_{mi}(t-\tau_m)
\varphi(m)dm} + X_{iA}(t) A(t)\\ \nonumber
 - X_{i}(t) W(t) +  X_{i}(t)I(t),
\end{eqnarray}
where $M_i$ can be substituted by $\sigma_i$, namely the surface gas density of the element $i$. In several models of chemical evolution it is customarily 
to use normalized variables which should be substituted to $M_i(t)$ or to 
$\sigma_i(t)$, such as for example:
\begin{equation}
G_i(t)= {\sigma_{i}(t) \over  \sigma_{tot}(t_G)}= {M_i(t) \over M_{tot}(t_G)}, 
\end{equation}
with $\sigma_{i}(t)=X_i(t) \sigma_{gas}(t)$, and $\sigma_{tot}(t_G)$ ($M_{tot}(t_G)$) being the total surface mass density (mass) at the present time $t_G$. 
The surface densities are more 
indicated for computing the chemical evolution of galactic disks, while for 
spheroids one can use the masses.
The quantity $X_i(t) = {\sigma_i(t) \over \sigma_{gas}(t)}= {M_i(t) \over M_{gas}(t)}$ represents
the abundance by mass of the element $i$
and by definition the summation over all the mass abundances
of the elements present in the gas
mixture is equal to unity.
These equations include the products from Type Ia SNe (third term on the right)
and the products of stars ending their lives as white dwarfs and core-collapse SNe (for a more extensive description of the equations see Matteucci, 2012).

\section{Some highlights}

In this section we will illustrate some examples where predictions from chemical evolution models are compared to observations.

\subsection{The chemical evolution of the Milky Way}
We will start with the Milky Way, which is the galaxy for which we possess the majority of information.
A good model for the chemical evolution of the Milky Way should reproduce several constraints, including the G-dwarf metallicity distribution, the [X/Fe] vs. [Fe/H]\footnote{[Fe/H]= $log(Fe/H)_* - log(Fe/H)_{\odot}$} relations (where X is a generic element from carbon up to the heaviest ones), abundance and gas gradients along the Galactic disk, among others.

In Figure 2 we show the predictions from Romano et al. (2010) concerning several chemical elements obtained by using different sets of stellar yields and compared to observation in stars. As one can see, the agreement is good for some chemical species whereas for others the agreement is still very poor. The reason for that probably resides in the uncertainties still existing in the theoretical stellar yields. The shapes of the [X/Fe] vs. [Fe/H] relation can be successfully interpreted as due to the {\it time-delay model}, namely the fact that elements such as C, N and Fe are mainly produced by long living stars, while others such as $\alpha$-elements (O, Mg, Si, Ca) are produced by short living stars. In particular, Fe is mainly produced by Type Ia SNe and only a small fraction of it originates in core-collapse (CC) SNe. Type I a SNe explode on longer timescales than core-collapse SNe and therefore ratios such as [$\alpha$/Fe] , where $\alpha$ indicates $\alpha$-elements which are mainly produced in CC SNe, can be used as cosmic clocks. The higher than solar value of the [$\alpha$/Fe] ratios for low [Fe/H] values,  is then due to the CC SNe which restore the $\alpha$-elements on short timescales. When Type Ia SNe, originating from CO white dwarfs which have longer lifetimes (from 30 Myr up to a Hubble time), start restoring the bulk of Fe, then the  [$\alpha$/Fe] ratios start decreasing. By means of the time-delay model we can interpret any abundace ratio.
\begin{figure}[h]
\includegraphics[width=20pc]{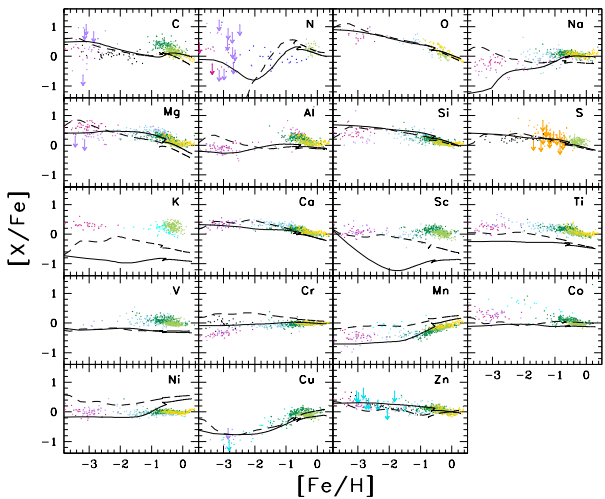}\hspace{2pc}%
\begin{minipage}[b]{14pc}\caption{Comparison between model predictions and observatioons. The two sets of models refer to different stellar yields: the dashed lines refer to the yields of  Woosley \& Weaver (1995) for massive stars and van den Hoek and Groenewegen (1997) for low and intermediate mass stars,
whereas the continuous lines represent the yields of Kobayashi et al. (2006) for massive stars, Karakas (2010) for low and intermediate masses  and for CNO elements the yields of the Geneva group with mass loss and rotation. Figure from Romano et al. (2010).}
\end{minipage}
\end{figure}

The model underlying the predictions of Figure 2 is an updated version of the two-infall model by Chiappini et al. (1997): this model assumes that the Milky Way formed by means of two main gas accretion episodes, one during which the halo and thick disk formed and another during which the thin disk formed on much longer timescales.
\subsection{The Li-problem}
In Figure 3 we show an illustration of the so-called Li-problem. In particular, we report the abundance of $^{7}Li$ (A(Li) = 12 +log(Li/H) versus [Fe/H]). The data sources are indicated inside the Figure. The upper envelope of the data should represent the evolution of the abundance of $^{7}Li$ in the ISM since the stars in the upper envelope should exhibit the Li abundance present in the gas when they formed. The stars below should instead have consumed the original Li which is very fragile and tends to react with protons to form  $^{4}He$ when the temperature is $>2.8 \cdot 10^{6}$ K. According to this interpretation of the diagram, the younger stars with higher [Fe/H] should reflect the fact that the Li abundance has increased with galactic lifetime owing to Li stellar production by stars. Stars which can create, preserve and eject Li into the ISM are: low and intermediate mass stars during the asymptotic giant branch (AGB) phase, the CC  SNe, cosmic rays and perhaps novae (see Izzo et al. 2015 for the possible detection of Li in a nova).
The Li problem arises from the fact that both WMAP and Planck experiments have derived a primordial $^{7}Li$ abundance higher than the abundance found in the upper envelope of the Galactic halo stars (those with [Fe/H]$<$ -1.0 in Figure 3). This could mean that the oldest Galactic stars have consumed their primordial Li and roughly by the same amount for stars in the range -3.0 ---  -1.0 in [Fe/H] (the so-called Spite plateau). The lines in Figure 3 represent the predictions of a chemical evolution model (the two-infall model) including SNe, novae, cosmic rays and AGB stars as Li producers and starting from the primordial Li abundance deduced either by WMAP or by the value of the Spite plateau. There is not yet a solution for the Li-problem. 

\begin{figure}[h]
\includegraphics[width=20pc]{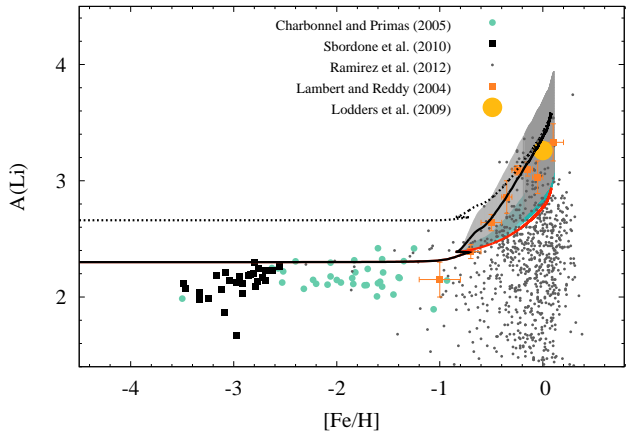}\hspace{2pc}%
\begin{minipage}[b]{14pc}\caption{A(Li) vs. [Fe/H] for solar vicinity stars and meteorites (symbols; see legend) compared to the predictions of chemical evolution
models (lines and colored areas). The chemical evolution model is from 
Romano et al. (2010).The black line represents all the stellar Li producers, as described in the text. The red line is the same model but without novae. Note that the dotted theoretical curve starts with
the primordial Li value from WMAP which lies well above the Li abundances of halo stars.Figure from Izzo et al. (2015).}
\end{minipage}
\end{figure}

\subsection{The Ultra-Faint Dwarfs (UFDs)} 
In Figure 4 we show the model predictions and data for the UFD galaxy Hercules, a small (with stellar mass of $\sim 10^{7}M_{\odot}$) satellite of the Milky Way.
The data refer to the [Ca/Fe] ratios measured in the stars of this galaxy and include also the data for Ca in the Milky Way and dwarf spheroidal stars. The data show that the Milky Way and Hercules evolved in a different way, since the [Ca/Fe] ratios are lower at the same  [Fe/H] in the UFD galaxy, relative to the Milky Way. The models, computed with an efficiency of star formation one thousand times lower than assumed for the Milky Way can well reproduce the trend. The model for Hercules includes also a strong galactic wind proportional to the SFR. Because of the already mentioned time-delay model, the Fe delayed production coupled with a regime of slow star formation predict that the [Ca/Fe] ratios are lower. at the same [Fe/H], since when the Type Ia SNe start to be important in producing Fe, the Ca and the other metals produced by CC SNe have not yet attained the same abundances as in the Miilky Way which evolves with a much faster star formation.
\begin{figure}[h]
\includegraphics[width=14pc]{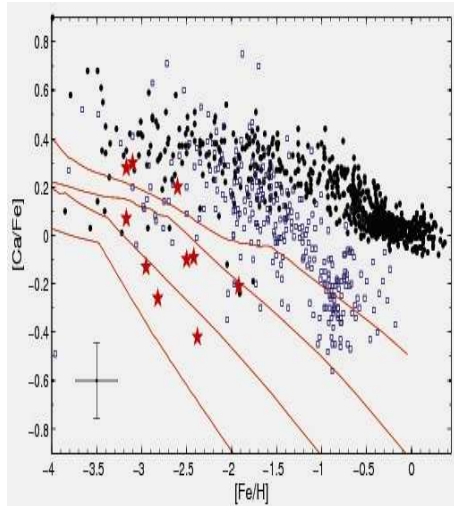}\hspace{2pc}%
\begin{minipage}[b]{14pc}\caption{Theoretical and observed [Ca/Fe] ratios as functions of [Fe/H]. The red stars represent abundances in stars of Hercules, while the blue points represent dwarf spheroidals and black points the Milky Way. The red lines are models assuming a star formation efficiency ($\nu$) varying in the range 0.002-0.008 $Gyr^{-1}$. Figure from Koch et al. (2012).}
\end{minipage}
\end{figure}
Therefore, in order to interpret the abundance patterns in galaxies we should 
consider the history of star formation, since it affects strongly the shape of the [X/Fe] vs. [Fe/H] relations. In this way, we can reconstruct the history of stars formation in galaxies just by looking at their abundances. This approach is known as {\it astroarchaeological approach}. From Figure 4 it apprears that the dwarf spheroidals have a higher star formation efficiency than the UFDs but a lower one than the Milky Way. From this Figure it is difficult to conclude that stars of the UFDs could have been the building blocks of the Galactic halo, but more data on UFDs are necessary before drawing firm conclusions on this point.

\subsection{Acknowledgments}
I acknowledge financial support from the PRIN2010-2011 project ``The Chemical and Dynamical Evolution of the Milky Way and Local Group Galaxies'', prot.2010LY5N2T.

\end{document}